\documentclass
[preprint,prd,onecolumn,showpacs,showkeys,nobibnotes,nofootinbib,titlepage]{revtex4}%
\usepackage{amsfonts}
\usepackage{amsmath}
\usepackage{amssymb}
\usepackage{graphicx}
\usepackage{float}
\usepackage{rotating}%
\providecommand{\U}[1]{\protect\rule{.1in}{.1in}}
\ifx\pdfoutput\relax\let\pdfoutput=\undefined\fi
\newcount\msipdfoutput
\ifx\pdfoutput\undefined\else
\ifcase\pdfoutput\else
\ifx\paperwidth\undefined\else
\ifdim\paperheight=0pt\relax\else\pdfpageheight\paperheight\fi
\ifdim\paperwidth=0pt\relax\else\pdfpagewidth\paperwidth\fi
\fi\fi\fi
\begin{document}
\preprint{ }
\title[Conformal Gravity Warp Drive]{Conformal Gravity and the Alcubierre Warp Drive Metric}
\author{Gabriele U. Varieschi}
\affiliation{Department of Physics, Loyola Marymount University - Los Angeles, CA 90045,
USA\footnote{Email: Gabriele.Varieschi@lmu.edu}}
\eid{ }
\author{Zily Burstein}
\affiliation{Department of Physics, Loyola Marymount University - Los Angeles, CA 90045,
USA\footnote{Email: zburstei@lion.lmu.edu}}
\author{}
\affiliation{}

\begin{abstract}
We present an analysis of the classic Alcubierre metric based on conformal
gravity, rather than standard general relativity. The main characteristics of
the resulting warp drive remain the same as in the original study by
Alcubierre, namely that effective super-luminal motion is a viable outcome of
the metric.

We show that for particular choices of the shaping function, the Alcubierre
metric in the context of conformal gravity does not violate the weak energy
condition, as was the case of the original solution. In particular, the
resulting warp drive does not require the use of exotic matter.

Therefore, if conformal gravity is a correct extension of general relativity,
super-luminal motion via an Alcubierre metric might be a realistic solution,
thus allowing faster-than-light interstellar travel.

\end{abstract}
\startpage{1}
\endpage{ }
\maketitle
\tableofcontents

\section{\label{sect:introduction}Introduction}

In 1994 M. Alcubierre introduced the so-called \textit{Warp Drive Metric
}(WDM), within the framework of General Relativity (GR), which allows in
principle for super-luminal motion, i.e., faster-than-light travel
\cite{Alcubierre:1994tu}. This super-luminal propulsion is achieved by
respectively expanding and contracting the space-time behind and in front of a
spaceship, while the spacecraft is left inside a locally flat region of
space-time, within the so-called \textit{warp bubble}.

In this way the spaceship can travel at arbitrarily high speeds, without
violating the laws of special and general relativity, or other known physical
laws. Furthermore, the spacecraft and its occupants would also be at rest in
flat space-time, thus immune from high accelerations and unaffected by special
relativistic effects, such as time dilation. Enormous tidal forces would only
be present near the edge of the warp bubble, which can be made large enough to
accommodate the volume occupied by the ship.

However, Alcubierre \cite{Alcubierre:1994tu}\ was also the first to point out
that this hypothetical solution of Einstein's equations of GR would violate
all three standard energy conditions (weak, dominant and strong; see
\cite{Hawking1}, \cite{Carroll:2004st}, and \cite{Wald1}\ for definitions). In
particular, the violation of the weak energy condition (WEC) implies that
negative energy density is required to establish the Alcubierre WDM, thus
making it practically impossible to achieve this type of super-luminal motion,
unless large quantities of exotic matter (i.e., with negative energy density)
can be created. Since our current knowledge of this type of exotic matter is
limited to some special effects in quantum field theory (such as the Casimir
effect), it is unlikely that the Alcubierre WDM can be practically established
within the framework of General Relativity.

Following Alcubierre's seminal paper, many other studies appeared in the
literature, either proposing alternatives to the original warp drive mechanism
(\cite{Krasnikov:1995ad}, \cite{Natario:2001tk}) or refining and analyzing in
more detail the original idea (\cite{Olum:1998mu}, \cite{VanDenBroeck:1999sn},
\cite{Clark:1999zn}, \cite{GonzalezDiaz:1999db}, \cite{Pfenning:1997wh},
\cite{2003GReGr..35.2025W}, \cite{Lobo:2004wq}, \cite{GonzalezDiaz:2007zza},
\cite{GonzalezDiaz:2007zzb}, \cite{Finazzi:2009jb}, \cite{Barcelo:2010pu},
\cite{Muller:2011fa}, \cite{McMonigal:2012ey}). However, all these studies
were conducted using standard GR and could not avoid the violation of the WEC,
meaning that some exotic matter would always be required for faster-than-light
travel. Similar issues also exist in other well-known GR solutions for
super-luminal motion, such as space-time wormholes \cite{Hartle1}.

Einstein's General Relativity and the related \textquotedblleft Standard
Model\textquotedblright\ of Cosmology have been highly successful in
describing our Universe, from the Solar System up to the largest cosmological
scales, but recently these theories have also led to a profound crisis in our
understanding of its ultimate composition. From the original discovery of the
expansion of the Universe, which resulted in standard Big Bang Cosmology,
scientists have progressed a long way towards our current picture, in which
the contents of the Universe are today described in terms of two main
components, dark matter (DM) and dark energy (DE), accounting for most of the
observed Universe, with ordinary matter just playing a minor role.

Since there is no evidence available yet as to the real nature of dark matter
and dark energy, alternative gravitational and cosmological theories are being
developed, in addition to standard explanations of dark matter/dark energy
invoking the existence of exotic new particles also yet to be discovered. In
line with these possible new theoretical ideas, \textit{Conformal Gravity}
(CG) has emerged as a non-standard extension of Einstein's GR, based on a
possible symmetry of the Universe: the conformal symmetry, i.e., the
invariability of the space-time fabric under local \textquotedblleft
stretching\textquotedblright\ of the metric (for reviews see
\cite{Mannheim:2005bfa}, \cite{Varieschi:2008fc}). This alternative theory has
been re-introduced in recent years (following the original work by H. Weyl
\cite{Weyl:1918aa}, \cite{Weyl:1918ib}, \cite{Weyl:1919fi}), leading to
cosmological models which do not require the existence of DM and DE
(\cite{Mannheim:2011ds}, \cite{Mannheim:2010ti}, \cite{Mannheim:2010xw},
\cite{O'Brien:2011wg}, \cite{Mannheim:2007ug}, \cite{Varieschi:2008va},
\cite{Diaferio:2011kc}).

In view of a possible extension of Einstein's General Relativity into
Conformal Gravity, in this paper we have re-considered the Alcubierre WDM,
basing it on CG rather than standard GR. In Sect. \ref{sect:conformal_gravity}
we review the fundamental principles of CG and the calculation of the
stress-energy tensor in this gravitational theory. In Sect.
\ref{sect:alcubierre_metric} we consider the Alcubierre metric in CG and
compute the energy density for different shaping functions of the metric.

In particular, we will show that for certain shaping functions, the Alcubierre
metric in the context of Conformal Gravity does not violate the weak energy
condition, as was the case of the original solution. This analysis continues
in Sect. \ref{sect:energy_conditions}, where we study other energy conditions
and estimate the total energy required for this CG warp drive. Finally, in
Sect. \ref{sect:conclusions}, we conclude that if CG is a correct extension of
GR, super-luminal motion via an Alcubierre metric might be a realistic
possibility, thus enabling faster-than-light interstellar travel without
requiring exotic matter.

\section{\label{sect:conformal_gravity}Conformal Gravity and the stress-energy
tensor}

H. Weyl in 1918 (\cite{Weyl:1918aa}, \cite{Weyl:1918ib}, \cite{Weyl:1919fi})
developed the \textquotedblleft conformal\textquotedblright\ generalization of
Einstein's relativity by introducing the conformal (or Weyl) tensor, a special
combination of the Riemann tensor $R_{\lambda\mu\nu\kappa}$, the Ricci tensor
$R_{\mu\nu}=R^{\lambda}{}_{\mu\lambda\nu}$ and the curvature (or Ricci) scalar
$R=R^{\mu}{}_{\mu}$ \cite{Weinberg}:%

\begin{equation}
C_{\lambda\mu\nu\kappa}=R_{\lambda\mu\nu\kappa}-\frac{1}{2}(g_{\lambda\nu
}R_{\mu\kappa}-g_{\lambda\kappa}R_{\mu\nu}-g_{\mu\nu}R_{\lambda\kappa}%
+g_{\mu\kappa}R_{\lambda\nu})+\frac{1}{6}R\ (g_{\lambda\nu}g_{\mu\kappa
}-g_{\lambda\kappa}g_{\mu\nu}), \label{eqn2.1}%
\end{equation}
where $C^{\lambda}{}_{\mu\lambda\nu}(x)$ is invariant under the local
transformation of the metric:%

\begin{equation}
g_{\mu\nu}(x)\rightarrow\widehat{g}_{\mu\nu}(x)=e^{2\alpha(x)}g_{\mu\nu
}(x)=\Omega^{2}(x)g_{\mu\nu}(x). \label{eqn2.2}%
\end{equation}
The factor $\Omega(x)=e^{\alpha(x)}$ determines the amount of local
\textquotedblleft stretching\textquotedblright\ of the geometry, hence the
name \textquotedblleft conformal\textquotedblright\ for a theory invariant
under all local stretchings of the space-time (see \cite{Varieschi:2008fc} and
references therein for more details).

This conformally invariant generalization of GR was found to be a fourth-order
theory, as opposed to the standard second-order General Relativity, since the
field equations originating from a conformally invariant Lagrangian contain
derivatives up to the fourth order of the metric, with respect to the
space-time coordinates. Following work done by R. Bach \cite{Bach:1921}, C.
Lanczos \cite{Lanczos:1938} and others, CG was ultimately based on the Weyl or
conformal action:\footnote{In this paper we adopt a metric signature (-,+,+,+)
and we follow the sign conventions of Weinberg \cite{Weinberg}. In this
section we will leave fundamental constants, such as $c$ and $G$, in all
equations, but later we will use geometrized units ($c=1$, $G=1$), or c.g.s.
units when needed.}%

\begin{equation}
I_{W}=-\alpha_{g}\int d^{4}x\ (-g)^{1/2}\ C_{\lambda\mu\nu\kappa}%
\ C^{\lambda\mu\nu\kappa}, \label{eqn2.3}%
\end{equation}
or on the following equivalent expression, differing from the previous one
only by a topological invariant:%
\begin{equation}
I_{W}=-2\alpha_{g}\int d^{4}x\ (-g)^{1/2}\ \left(  R_{\mu\kappa}R^{\mu\kappa
}-\frac{1}{3}R^{2}\right)  , \label{eqn2.4}%
\end{equation}
where $g\equiv\det(g_{\mu\nu})$ and $\alpha_{g}$ is the gravitational coupling
constant of Conformal Gravity (see \cite{Mannheim:2005bf},
\cite{Schimming:2004yx}, \cite{Mannheim:1988dj}, \cite{Kazanas:1988qa}%
).\footnote{In these cited papers, $\alpha_{g}$ is considered a dimensionless
constant\ by using natural units. Working with c.g.s. units, we can assign
dimensions of an action to the constant $\alpha_{g}$ so that the
dimensionality of Eq. (\ref{eqn2.5}) will be correct.} Under the conformal
transformation in Eq. (\ref{eqn2.2}), the Weyl tensor transforms as
$C_{\lambda\mu\nu\kappa}\rightarrow\widehat{C}_{\lambda\mu\nu\kappa
}=e^{2\alpha(x)}C_{\lambda\mu\nu\kappa}=\Omega^{2}(x)C_{\lambda\mu\nu\kappa}$,
while the conformal action $I_{W}$ is locally conformally invariant, the only
general coordinate scalar action with such properties.

R. Bach \cite{Bach:1921} introduced the gravitational field equations in the
presence of a stress-energy tensor\footnote{We follow here the convention
\cite{Mannheim:2005bf}\ of introducing the stress-energy tensor $T_{\mu\nu}$
so that the quantity $cT_{00}$ has the dimensions of an energy density.}
$T_{\mu\nu}$ :%

\begin{equation}
W_{\mu\nu}=\frac{1}{4\alpha_{g}}\ T_{\mu\nu} \label{eqn2.5}%
\end{equation}
as opposed to Einstein's standard\ equations,%

\begin{equation}
G_{\mu\nu}=R_{\mu\nu}-\frac{1}{2}g_{\mu\nu}\ R=-\frac{8\pi G}{c^{3}}%
\ T_{\mu\nu}, \label{eqn2.6}%
\end{equation}
where the \textquotedblleft Bach tensor\textquotedblright\ $W_{\mu\nu}$
\cite{Bach:1921}\ is the equivalent in CG of the Einstein curvature tensor
$G_{\mu\nu}$ on the left-hand side of Eq. (\ref{eqn2.6}).

$W_{\mu\nu}$ has a very complex structure and can be defined in a compact way
as \cite{Schmidt:1984bg}:%

\begin{equation}
W_{\mu\nu}=2C^{\alpha}{}_{\mu\nu}{}^{\beta}{}_{;\beta;\alpha}+C^{\alpha}%
{}_{\mu\nu}{}^{\beta}\ R_{\beta\alpha}, \label{eqn2.7}%
\end{equation}
or in an expanded form as (\cite{Mannheim:1988dj}, \cite{Wood:2001ve}):%

\begin{align}
W_{\mu\nu}  &  =-\frac{1}{6}g_{\mu\nu}\ R^{;\lambda}{}_{;\lambda}+\frac{2}%
{3}R_{;\mu;\nu}+R_{\mu\nu}{}^{;\lambda}{}_{;\lambda}-R_{\mu}{}^{\lambda}%
{}_{;\nu;\lambda}-R_{\nu}{}^{\lambda}{}_{;\mu;\lambda}+\frac{2}{3}R\ R_{\mu
\nu}\label{eqn2.8}\\
&  -2R_{\mu}{}^{\lambda}\ R_{\lambda\nu}+\frac{1}{2}g_{\mu\nu}\ R_{\lambda
\rho}\ R^{\lambda\rho}-\frac{1}{6}g_{\mu\nu}\ R^{2},\nonumber
\end{align}
involving derivatives up to the fourth order of the metric with respect to
space-time coordinates.

Therefore, in Conformal Gravity, the stress-energy tensor is computed by
combining together Eqs. (\ref{eqn2.5}) and (\ref{eqn2.8}):%

\begin{align}
T_{\mu\nu}  &  =4\alpha_{g}\ W_{\mu\nu}=4\alpha_{g}\ (-\frac{1}{6}g_{\mu\nu
}\ R^{;\lambda}{}_{;\lambda}+\frac{2}{3}R_{;\mu;\nu}+R_{\mu\nu}{}^{;\lambda}%
{}_{;\lambda}-R_{\mu}{}^{\lambda}{}_{;\nu;\lambda}-R_{\nu}{}^{\lambda}{}%
_{;\mu;\lambda}\label{eqn2.9}\\
&  +\frac{2}{3}R\ R_{\mu\nu}-2R_{\mu}{}^{\lambda}\ R_{\lambda\nu}+\frac{1}%
{2}g_{\mu\nu}\ R_{\lambda\rho}\ R^{\lambda\rho}-\frac{1}{6}g_{\mu\nu}%
\ R^{2}).\nonumber
\end{align}
This form of the tensor will be used in the following sections, in connection
to the Alcubierre metric, to compute the energy density and other relevant quantities.

For this purpose, we have developed a special Mathematica program which
enables us to compute all the tensor quantities of both GR and CG, for any
given metric. In particular, this program can compute the conformal tensor
$C_{\lambda\mu\nu\kappa}$\ in Eq. (\ref{eqn2.1}), the Bach tensor\ $W_{\mu\nu
}$ in Eq. (\ref{eqn2.8}), or the stress-energy tensor $T_{\mu\nu}$\ in Eq.
(\ref{eqn2.9}) by performing all the necessary covariant derivatives. Given
the complexity of these types of computations, we have tested extensively our
program against the results for $C_{\lambda\mu\nu\kappa}$\ and $W_{\mu\nu}$
computed by Mannheim et al. in Ref. \cite{Kazanas:1988qa} for different
metrics, obtaining a perfect agreement.

\section{\label{sect:alcubierre_metric}Alcubierre metric, shaping functions
and the weak energy condition}

The original Alcubierre metric \cite{Alcubierre:1994tu} considered a spaceship
traveling along the x-axis, with motion described by a function $x_{s}(t)$ and
spaceship velocity $v_{s}(t)=\frac{dx_{s}(t)}{dt}$. Using the 3+1 formalism of
GR, the metric was written in Cartesian coordinates as ($c=1$):%

\begin{equation}
ds^{2}=-dt^{2}+\left[  dx-v_{s}(t)\ f(r_{s})\ dt\right]  ^{2}+dy^{2}+dz^{2},
\label{eqn3.1}%
\end{equation}
where $r_{s}$ is the distance from the spaceship position:%

\begin{equation}
r_{s}(t)=\sqrt{\left[  x-x_{s}(t)\right]  ^{2}+y^{2}+z^{2}}, \label{eqn3.2}%
\end{equation}
and $f(r_{s})$ is a \textquotedblleft form function\textquotedblright\ or
\textquotedblleft shaping function\textquotedblright\ which needs to have
values $f=1$ and $f=0$ respectively inside and outside the warp bubble, while
it can have an arbitrary shape in the transition region of the warp bubble itself.

The original shaping function used by Alcubierre was:%

\begin{equation}
f(r_{s})=\frac{\tanh\left[  \sigma\left(  r_{s}+R\right)  \right]
-\tanh\left[  \sigma\left(  r_{s}-R\right)  \right]  }{2\tanh\left(  \sigma
R\right)  }, \label{eqn3.3}%
\end{equation}
where $R>0$ basically indicates the radius of the spherical warp bubble, while
$\sigma>0$ relates to the bubble thickness, which decreases with increasing
values of $\sigma$. In the following, we will refer to the function in Eq.
(\ref{eqn3.3}) as the \textquotedblleft Alcubierre shaping
function\textquotedblright\ (ASF).

We will show that the particular form of the shaping function can play an
important role in the energy conditions for the WDM. In our analysis we tested
several different functions obeying the general requirements for $f$ outlined
above. In addition to the Alcubierre function above, in this paper we will
also use the following:%

\begin{equation}
f(r_{s})=%
\begin{Bmatrix}
1-\left(  \frac{r_{s}}{R}\right)  ^{m} & ;\ r_{s}<R\\
0 & ;\ r_{s}>R
\end{Bmatrix}
, \label{eqn3.4}%
\end{equation}
where $m$ is a positive integer. Since this particular function for $m=4$ is
used by J. Hartle to illustrate the warp drive in his textbook \cite{Hartle1},
we will refer to the function in Eq. (\ref{eqn3.4}) as the \textquotedblleft
Hartle shaping function\textquotedblright\ (HSF).

The top panels in Fig. 1 illustrate the differences between the Alcubierre
shaping function (top left panel, for $\sigma=8$ as used originally by
Alcubierre) and the Hartle shaping function (top right panel, for $m=4$). All
functions in this figure are computed for a fixed value of $R=1$ and at time
$t=0$, when the spaceship is located at the origin. All quantities shown in
the different panels are plotted as a function of the $x$ coordinate of the
spaceship motion and of the transverse cylindrical coordinate $\rho
=\sqrt{y^{2}+z^{2}}$. Similar coordinates will also be used in the other
figures.\footnote{The cylindrical coordinate $\rho=\sqrt{y^{2}+z^{2}}$ should
be considered as non-negative and all quantities in the figures plotted only
for $\rho\geq0$. However, for illustrative purposes and also to follow similar
figures in the literature (such as those in \cite{Alcubierre:1994tu},
\cite{Pfenning:1997wh}, \cite{2003GReGr..35.2025W}, etc.), we decided to let
$\rho$ run on negative values in all figures, except in the last one, where we
restrict $\rho\geq0$ for a correct energy calculation.}

The expansion/contraction function $\theta$\ of the volume elements behind/in
front of the spaceship was also computed by Alcubierre as
\cite{Alcubierre:1994tu}:%

\begin{equation}
\theta=v_{s}(t)\frac{\left[  x-x_{s}(t)\right]  }{r_{s}}\frac{df(r_{s}%
)}{dr_{s}}, \label{eqn3.5}%
\end{equation}
and is illustrated for $v_{s}=1$\ (in geometrized units, i.e., $v_{s}=c$ in
traditional units) in the second row of Fig. 1, for the two different shaping
functions. Again, the choice of the parameters $R$, $\sigma$, and $m$ is the
same as in the top panels in the figure. The expansion $\theta$ for the ASF is
the same as Fig. 1 in Ref. \cite{Alcubierre:1994tu}, while the corresponding
$\theta$ for the HSF is slightly different, but still shows expansion of the
normal volume elements behind the spaceship and contraction in front of it.

The weak energy condition (\cite{Hawking1}, \cite{Carroll:2004st},
\cite{Wald1}) requires that $T_{\mu\nu}t^{\mu}t^{\nu}\geq0$ for all timelike
vectors $t^{\mu}$. Alcubierre has also shown that for the Eulerian observers
in the warp drive metric, and for their 4-velocity $n_{\mu}$, the following
relation holds \cite{Alcubierre:1994tu}:%

\begin{equation}
T^{\mu\nu}n_{\mu}n_{\nu}=T^{00}=-\frac{1}{8\pi}\frac{v_{s}^{2}\rho^{2}}%
{4r_{s}^{2}}\left(  \frac{df}{dr_{s}}\right)  ^{2}, \label{eqn3.6}%
\end{equation}
which implies that the energy density $T^{00}$ is negative everywhere for any
choice of the shaping function $f$ and, therefore, the WEC is violated (also
the dominant energy condition -DEC- and the strong energy condition -SEC- are
violated in the analysis based on GR \cite{Alcubierre:1994tu}).

This violation of the WEC in GR\ is illustrated in the third row of Fig. 1,
where $T^{00}$ is calculated using Eq. (\ref{eqn3.6}) for both shaping
functions. Although the results in the two panels are slightly different, they
obviously show negative energy densities and therefore a complete violation of
the WEC.

The situation is different if we compute the energy density $T^{00}$ in the
framework of CG, following Eq. (\ref{eqn2.9}), setting $\alpha_{g}=1$ for
simplicity, and using the completely contravariant form of the stress-energy
tensor, instead of the covariant one. As seen in the bottom row of Fig. 1, the
energy density in CG\ is completely different from the one calculated within
the framework of GR. In the bottom left panel $T^{00}$ is computed with the
ASF and the resulting function is in part positive and in part negative, thus
still violating the WEC.

However, the bottom right panel shows $T^{00}$ computed with the HSF and in
this case the energy density is completely non-negative, showing that the WEC
is verified and no exotic matter is needed to establish the warp drive. This
non-negative energy density plot in the bottom right panel of Fig. 1 is the
main result of our paper as it shows that - if CG is the correct extension of
GR - it might be possible to establish a warp drive without having to use
negative energy (mass), thus overcoming the main difficulty of the warp drive mechanism.

The explicit expression of $T^{00}$ in CG, computed with our Mathematica
program, is rather cumbersome and is reproduced in Eq. (\ref{eqnA.1}) of the
Appendix. Here we present just the graphical computation of $T^{00}$ in Fig.
1, or in the other figures in this paper. We have tested the validity of these
results by running the program in several different ways, including computing
the stress-energy tensor for a simplified three-dimensional Alcubierre metric
(coordinates $x$, $y$, $t$ only), always obtaining consistent results. The
differences in the two plots at the bottom of the figure can be attributed to
the different shaping functions and their derivatives up to the fourth order.
All these derivatives enter the complex expression of $T^{\mu\nu}$ in CG, as
in the master Eq. (\ref{eqn2.9}), and their interplay ultimately determines
the shape of $T^{00}$, or of the other components, in a way which is hard to
predict before the actual computation is performed.

In figure 2 we analyze the dependence of the energy density $T^{00}$ on the
spaceship velocity $v_{s}$. In this case we consider only the Hartle shaping
function for $m=4$ and $R=1$, and we compute $T^{00}$ in CG ($\alpha_{g}=1$)
for speeds ranging from the sub-luminal $v_{s}=0.25\ c$ to the super-luminal
$v_{s}=3.00\ c$ (for $v_{s}=1.00\ c$ we obtain the same function as in the
bottom right panel of Fig. 1). The shape of the energy density function is
about the same for speeds up to $v_{s}=1.50\ c$, although the function values
increase with speed. For higher velocities, the function develops two
\textquotedblleft downward lobes\textquotedblright\ which eventually become
negative for speeds $v_{s}\gtrsim2.50\ c$. This implies that the WEC is
verified for speeds up to $v_{s}\approx2.50\ c$, while at higher velocities
exotic matter would be required to sustain the warp drive.

This apparent \textquotedblleft speed limit\textquotedblright\ at about
$v_{s}\approx2.50\ c$ might be raised or overcome completely by adopting a
different shaping function, instead of the HSF used here, but this analysis
would go beyond the scope of this work. In any case, the results reported in
Fig. 2 show that a warp drive in CG with positive energy density is possible
for a wide range of spaceship velocities; therefore, if CG is the correct
extension of GR, the Alcubierre warp drive might be a viable mechanism for
super-luminal travel.

In Fig. 3 we present the other components of the stress-energy tensor. These
were computed with the same Mathematica program, following Eq. (\ref{eqn2.9})
with $\alpha_{g}=1$, leading to even more complex expressions than the one for
$T^{00}$ (we will omit to report these expressions for brevity). To simplify
the computation, we used the covariant components $T_{\mu\nu}$\ and adopted
cylindrical coordinates around the x-axis, $(t,x,\rho,\phi)\equiv(0,1,2,3)$,
instead of the Cartesian coordinates of the original Alcubierre metric. The
results in the different panels are labeled accordingly. We recall that the
stress-energy tensor is symmetric, $T_{\mu\nu}=T_{\nu\mu}$, and only the
non-zero components are illustrated in this figure, under similar conditions
used before (CG with HSF and parameters $m=4$, $R=1$, $v_{s}=1.00\ c$).

The shapes of the other components of $T^{\mu\nu}$ are more complex than the
one of $T^{00}$, but they can all be determined analytically, either in
covariant or contravariant form, using our program. If this exact form of the
stress-energy tensor could be established in the region surrounding the
spacecraft, warp drive motion would be feasible within the framework of
Conformal Gravity.

We also want to point out that we set the spaceship motion in the positive
direction of the x-axis (setting $v_{s}=+1.00\ c$), and this is reflected in
the symmetry, or lack thereof, of the components of $T_{\mu\nu}$. While some
components, such as $T^{00}$ (or $T_{00}$), $T_{01}=T_{10}$, $T_{11}$,
$T_{22}$, $T_{33}$, appear to be symmetric under the exchange $x\rightarrow
-x$, the other components, $T_{02}=T_{20}$ and $T_{12}=T_{21}$, are not
symmetric under this exchange and, therefore, these components must contain
information about the spaceship direction of motion. This argument overcomes
the objection addressed in Ref. \cite{2003GReGr..35.2025W} that since $T^{00}$
is symmetric about the $x_{s}=0$ plane, there is uncertainty in where the
space-time is expanded/contracted, thus making it impossible for the spaceship
to know in which direction of the x-axis, positive or negative, to move.
Rather, Fig. 3 shows that a \textquotedblleft bias\textquotedblright\ towards
one of the two possible directions is induced by some components of $T_{\mu
\nu}$.

Fig. 4 illustrates one last dependence of the energy density $T^{00}$ on the
parameters used. In this case, we set $R=1$, $v_{s}=1.00\ c$ and consider the
Hartle shaping function as in Eq. (\ref{eqn3.4}), while varying the\ integer
parameter $m$. In addition to our standard value, $m=4$, we have also tried
values from $m=2$ to $m=10$, as shown in the figure.

In general, increasing the $m$ value increases the internal volume of the warp
bubble, where space-time is flat, and therefore decreases the thickness of the
bubble wall where the space-time distortion takes place, in a way similar to
that of the $\sigma$ parameter in the original Alcubierre warp drive. However,
increasing $m$ also increases the energy required to establish the warp drive
(as we checked by integrating the functions in Fig. 4). Therefore, it appears
to be convenient to use a low value for this parameter. As shown in the
different panels of the figure, the first value, $m=2$, does not work since it
does not leave a flat space-time volume inside the bubble, while $m=3$ seems
to create a very small volume inside the bubble. Therefore, intermediate
values such as $m=4-6$ would appear to be more adequate to establish the warp
drive. Meanwhile, the solution for $m=10$ would require much more energy and
would not give any advantage, except reducing the thickness of the warp
bubble. We also checked that changing the value of $m$ does not have a strong
effect on the \textquotedblleft speed limit\textquotedblright\ of
$v_{s}\approx2.50\ c$, reported above for the case $m=4$. Thus, this value of
the parameter seems to be the most adequate for this type of solutions.

\section{\label{sect:energy_conditions}Other energy conditions and warp drive
energy estimate}

In the previous section we have discussed at length the weak energy condition
-WEC- for the Conformal Gravity Alcubierre warp drive. We have seen that, if
the Hartle shaping function is used, this condition is not violated for a wide
range of spaceship velocities, including super-luminal speeds. In this section
we will briefly analyze the other main energy conditions and estimate the
energy necessary to establish the warp drive in CG.

The dominant energy condition -DEC- is reported in the literature
(\cite{Hawking1}, \cite{Carroll:2004st}, \cite{Wald1})\ as $T^{00}%
\geq\left\vert T^{\mu\nu}\right\vert $ for any $\mu$, $\nu$, or equivalently
as assuming the WEC plus the additional condition that $T^{\mu\nu}t_{\mu}$ is
a non-spacelike vector, i.e., $T_{\mu\nu}T^{\nu}\,_{\lambda}t^{\mu}t^{\lambda
}\leq0$. It is easy to see that using as a vector $t_{\mu}$ the 4-velocity
$n_{\mu}$ of the Eulerian observers \cite{Alcubierre:1994tu}, the previous
condition for the DEC becomes $T^{0}\,_{\lambda}T^{0\lambda}\leq0$.

Figure 5 illustrates the violation of the DEC for our standard solution (AWD
with HSF and $m=4$, $R=1$, $v_{s}=1.00\ c$). The plotted function
$T^{0}\,_{\lambda}T^{0\lambda}$ is not negative everywhere, as required by the
DEC, but shows a violation for the central portion of the warp bubble. Even if
this energy condition appears to be violated, this does not notably affect the
feasibility of our CG warp drive. We recall that the DEC is usually related to
the standard perfect fluid stress-energy tensor, $T_{\mu\nu}=(\rho+p)U_{\mu
}U_{\nu}+pg_{\mu\nu}$, where here $\rho$ and $p$ are the fluid density and
pressure, while $U_{\mu}$ is the fluid 4-velocity. In this context the DEC
requires $\rho\geq\left\vert p\right\vert $, but this condition is not
required in general by all classical forms of matter \cite{Carroll:2004st};
therefore, its violation in our case is not particularly significant.

On the contrary, our standard solution also verifies the strong energy
condition -SEC- which states that $T_{\mu\nu}t^{\mu}t^{\nu}\geq\frac{1}%
{2}T^{\lambda}\,_{\lambda}t^{\sigma}t_{\sigma}$ for all timelike vectors
$t^{\mu}$. Again, using the Eulerian 4-velocity vector in place of $t^{\mu}$,
the previous condition is equivalent to $T^{00}+\frac{1}{2}T^{\lambda
}\,_{\lambda}\geq0$. Since the scalar $T^{\lambda}\,_{\lambda}$ is identically
zero for all our solutions, as checked using our Mathematica program, the SEC
is equivalent to $T^{00}\geq0$, which is the WEC already verified in Sect.
\ref{sect:alcubierre_metric}.

Finally, we want to estimate the energy necessary to establish our CG warp
drive, under reasonable conditions. For this purpose, in Fig. 6 we computed
once again the energy density $T^{00}$ for our AWD with the Hartle shaping
function ($\alpha_{g}=1$, $m=4$, $v_{s}=1.00\ c$), but this time for
$R=10000\ cm=100\ m$, a reasonable radius for a warp bubble enclosing our spaceship.

Figure 6 illustrates this solution, plotted only for $\rho=\sqrt{y^{2}+z^{2}%
}\geq0$, as this is the correct interval for the transverse coordinate $\rho$.
The cylindrical symmetry of this solution can also be better appreciated in
this type of plot. We then followed the procedure outlined in Refs.
\cite{VanDenBroeck:1999sn}, \cite{Pfenning:1997wh}, to integrate the local
energy density over the proper volume, in cylindrical coordinates at time
$t=0$ over all space, obtaining the total energy $E$:%

\begin{equation}
E=\alpha_{g}\ c\int d^{3}x\sqrt{\left\vert g\right\vert }\ T^{00}=\left(
1.86\times10^{10}\ \alpha_{g}\right)  \
%TCIMACRO{\unit{erg}}%
%BeginExpansion
\operatorname{erg}%
%EndExpansion
, \label{eqn4.1}%
\end{equation}
where $g=Det\ \left\vert g_{ij}\right\vert $ is the determinant of the spatial
metric on the constant time hypersurface. Since we assume that the spaceship
is traveling at constant velocity, $v_{s}=1.00\ c$, the total energy is also
constant with time. In the last equation, we reinstated a factor of $c$ to
obtain the correct dimensions (see footnote before Eq. (\ref{eqn2.5})) and
also inserted an overall multiplicative factor $\alpha_{g}$, which corresponds
to the conformal gravity coupling constant in Eq. (\ref{eqn2.9}). This factor
is necessary since our computation of $T^{00}$ in Fig. 6\ was done assuming
$\alpha_{g}=1$.

Therefore, we need to know the CG value for $\alpha_{g}$ in order to complete
our energy estimation. Unfortunately, the value of this coupling constant is
not well determined yet. The only value in the literature is reported by P.
Mannheim \cite{Mannheim:2007ug}:%

\begin{equation}
\alpha_{g}=3.29\times10^{82}\
%TCIMACRO{\unit{erg}}%
%BeginExpansion
\operatorname{erg}%
%EndExpansion%
%TCIMACRO{\unit{s}}%
%BeginExpansion
\operatorname{s}%
%EndExpansion
, \label{eqn4.2}%
\end{equation}
since this coupling constant has the dimensions of action. Inserting this
value for $\alpha_{g}$ in Eq. (\ref{eqn4.1}), we obtain the energy estimate:%

\begin{equation}
E=6.12\times10^{92}%
%TCIMACRO{\unit{erg}}%
%BeginExpansion
\operatorname{erg}%
%EndExpansion
=6.81\times10^{71}%
%TCIMACRO{\unit{g}}%
%BeginExpansion
\operatorname{g}%
%EndExpansion
=3.42\times10^{38}M_{\odot}, \label{eqn4.3}%
\end{equation}
where we also converted this energy into equivalent mass and compared our
result with the Solar mass $M_{\odot}=1.99\times10^{33}%
%TCIMACRO{\unit{g}}%
%BeginExpansion
\operatorname{g}%
%EndExpansion
$.

The estimate in Eq. (\ref{eqn4.3}) would imply that an enormous amount of
(standard) mass-energy is needed to establish our warp drive at a velocity
equal to the speed of light, with a reasonable size for the warp bubble.
However, the CG value of $\alpha_{g}$ is not well-established, since the
number in Eq. (\ref{eqn4.2}) represents only an estimate of the macroscopic
value of this coupling constant. This does not need to be the same as the
microscopic value associated with the fundamental theory, which could be
reduced by a factor of $N$, where $N$ could be the number of occupied baryonic
states in a galaxy ($N\sim10^{68}$), or possibly the number of baryons in the
Universe ($N\sim10^{80}$) \cite{Mannheim:2012}.

Therefore, our estimate could be reduced by many orders of magnitude.
Moreover, the energy necessary to establish the warp drive might also be
decreased by using a more efficient shaping function, an analysis which we
leave for a future study on the subject.

\section{\label{sect:conclusions}Conclusions}

In this paper we have analyzed in detail the Alcubierre warp drive mechanism
within the framework of Conformal Gravity. We have seen that a particular
choice of the shaping function (Hartle shaping function, instead of the
original Alcubierre one) can overcome the main limitation of the AWD in
standard General Relativity, namely the violation of the weak energy condition.

In fact, we have shown that for a wide range of spaceship velocities, the CG
solutions do not violate the WEC, and, therefore, the AWD mechanism might be
viable, if CG is the correct extension of the current gravitational theories.
All the components of the stress-energy tensor can be analytically calculated,
using a Mathematica program based on Conformal Gravity. Thus, a warp drive
can, at least in principle, be fully established following our computations.

We have also checked two other main energy conditions: the SEC is always
verified, while the DEC is violated, at least in the case we considered.
Finally, we estimated the energy needed to establish a reasonable warp drive
at the speed of light. This energy depends critically on the value of
$\alpha_{g}$, the CG coupling constant, which is not well known. Therefore,
this estimate will need to be refined in future studies.

\section{Appendix: energy density expression in Conformal Gravity}

We present here the expression for the energy density $T^{00}$\ in Conformal
Gravity, computed using our Mathematica program. This is the general form of
$T^{00}$\ for any shaping function $f[r_{s}$$]$ and its derivatives, up to the
fourth order. The energy density is a function of coordinates $t$, $x$,
$\rho=\sqrt{y^{2}+z^{2}}$, and, therefore, it has a cylindrical symmetry
around the x-axis. The distance $r_{s}$ is defined in Eq. (\ref{eqn3.2}),
$v_{s}$ is the spaceship velocity and $\alpha_{g}$ is the Conformal Gravity
coupling constant.%

\begin{align}
T^{00}  &  =\alpha_{g}\frac{\text{$v_{s}$}^{2}}{3\text{$r_{s}$}^{6}%
}(-4\text{$r_{s}v_{s}$}^{2}\rho^{2}\left(  6(x-\text{$v_{s}t$})^{2}+5\rho
^{2}\right)  (-1+f[\text{$r_{s}$}])f^{\prime}[\text{$r_{s}$}]^{3}%
\tag{A.1}\label{eqnA.1}\\
&  +4r_{s}^{4}v_{s}^{2}\left(  (x-\text{$v_{s}t$})^{2}+3\rho^{2}\right)
f^{\prime}[r_{s}]^{4}+f^{\prime}[r_{s}]^{2}(-24(x-\text{$v_{s}t$}%
)^{4}+3\left(  1+5\text{$v_{s}$}^{2}\right)  (x-\text{$v_{s}t$})^{2}\rho
^{2}\nonumber\\
&  +\left(  27+10\text{$v_{s}$}^{2}\right)  \rho^{4}+\text{$v_{s}$}^{2}%
(5\rho^{2}\left(  3(x-\text{$v_{s}t$})^{2}+2\rho^{2}\right)
(-2+f[\text{$r_{s}$}])f[\text{$r_{s}$}]-4\text{$r_{s}$}^{2}(x-\text{$v_{s}t$%
})^{2}\nonumber\\
&  \left(  4(x-\text{$v_{s}t$})^{2}+3\rho^{2}\right)  (-1+f[\text{$r_{s}$%
}])f^{\prime\prime}[\text{$r_{s}$}]))-2\text{$r_{s}$}f^{\prime}[\text{$r_{s}$%
}](16(x-\text{$v_{s}t$})^{2}-8\rho^{2}+(-60(x-\text{$v_{s}t$})^{4}\nonumber\\
&  +\left(  -97+5\text{$v_{s}$}^{2}\right)  (x-\text{$v_{s}t$})^{2}\rho
^{2}+\left(  -37+3\text{$v_{s}$}^{2}\right)  \rho^{4})f^{\prime\prime
}[\text{$r_{s}$}]+\text{$r_{s}($}4\left(  -4+\text{$v_{s}$}^{2}\right)
(x-\text{$v_{s}t$})^{4}\nonumber\\
&  +3\left(  -9+\text{$v_{s}$}^{2}\right)  (x-\text{$v_{s}t$})^{2}\rho
^{2}-11\rho^{4})f^{(3)}[\text{$r_{s}$}]+\text{$v_{s}$}^{2}f[\text{$r_{s}$%
}]^{2}(\rho^{2}\left(  5(x-\text{$v_{s}t$})^{2}+3\rho^{2}\right)
f^{\prime\prime}[\text{$r_{s}$}]\nonumber\\
&  +\text{$r_{s}$}(x-\text{$v_{s}t$})^{2}\left(  4(x-\text{$v_{s}t$}%
)^{2}+3\rho^{2}\right)  f^{(3)}[\text{$r_{s}$}])+2f[\text{$r_{s}$%
}](-8(x-\text{$v_{s}t$})^{2}+4\rho^{2}+\text{$v_{s}$}^{2}\nonumber\\
&  \left(  -\rho^{2}\left(  5(x-\text{$v_{s}t$})^{2}+3\rho^{2}\right)
f^{\prime\prime}[\text{$r_{s}$}]-\text{$r_{s}$}(x-\text{$v_{s}t$})^{2}\left(
4(x-\text{$v_{s}t$})^{2}+3\rho^{2}\right)  f^{(3)}[\text{$r_{s}$}]))\right)
\nonumber\\
&  +\text{$r_{s}$}^{2}(-16\left(  2(x-\text{$v_{s}t$})^{2}-\rho^{2}\right)
(-1+f[\text{$r_{s}$}])f^{\prime\prime}[\text{$r_{s}$}]+(4\left(
6+\text{$v_{s}$}^{2}\right)  (x-\text{$v_{s}t$})^{4}+\left(  43+3\text{$v_{s}%
$}^{2}\right) \nonumber\\
&  (x-\text{$v_{s}t$})^{2}\rho^{2}+19\rho^{4}+\text{$v_{s}$}^{2}%
(x-\text{$v_{s}t$})^{2}\left(  4(x-\text{$v_{s}t$})^{2}+3\rho^{2}\right)
(-2+f[\text{$r_{s}$}])f[\text{$r_{s}$}])f^{\prime\prime}[\text{$r_{s}$}%
]^{2}\nonumber\\
&  +8\text{$r_{s}$}(-1+f[\text{$r_{s}$}])\left(  \left(  2(x-\text{$v_{s}t$%
})^{2}+\rho^{2}\right)  f^{(3)}[\text{$r_{s}$}]+\text{$r_{s}$}(x-\text{$v_{s}%
t$})^{2}f^{(4)}[\text{$r_{s}$}]))\right) \nonumber
\end{align}

In the main part of our work, we used the Hartle shaping function in Eq.
(\ref{eqn3.4}), or more explicitly:%

\begin{equation}
f\left[  r_{s}\right]  =\left[  1-\left(  \frac{r_{s}}{R}\right)  ^{m}\right]
^{+}\equiv\frac{\left\vert 1-\left(  \frac{r_{s}}{R}\right)  ^{m}\right\vert
+\left[  1-\left(  \frac{r_{s}}{R}\right)  ^{m}\right]  }{2}, \tag{A.2}%
\label{eqnA.2}%
\end{equation}
where $\left[  ...\right]  ^{+}$ indicates the positive part of the function.
The derivatives of the HSF, up to the fourth order, were computed in terms of
the Heaviside step function $H(x)$ and the Dirac delta function $\delta(x)$,
also using the following relations for the derivatives of these special
functions: $\frac{d\left\vert x\right\vert }{dx}=sgn(x)=2H(x)-1$;
$\frac{dH(x)}{dx}=\delta(x)$; $x^{n}\frac{d^{n}\delta(x)}{dx^{n}}%
=(-1)^{n}\ n!\ \delta(x)$.%

\begin{equation}
f^{\prime}\left[  r_{s}\right]  =-\frac{m}{R}\left(  \frac{r_{s}}{R}\right)
^{m-1}H\left[  1-\left(  \frac{r_{s}}{R}\right)  ^{m}\right]  \tag{A.3}%
\label{eqnA.3}%
\end{equation}

\begin{equation}
f^{^{\prime\prime}}\left[  r_{s}\right]  =-\frac{m(m-1)}{R^{2}}\left(
\frac{r_{s}}{R}\right)  ^{m-2}H\left[  1-\left(  \frac{r_{s}}{R}\right)
^{m}\right]  +\frac{m^{2}}{R^{2}}\left(  \frac{r_{s}}{R}\right)  ^{2m-2}%
\delta\left[  1-\left(  \frac{r_{s}}{R}\right)  ^{m}\right]  \tag{A.4}%
\label{eqnA.4}%
\end{equation}

\begin{align}
f^{(3)}\left[  r_{s}\right]   &  =-\frac{m(m-1)(m-2)}{R^{3}}\left(
\frac{r_{s}}{R}\right)  ^{m-3}H\left[  1-\left(  \frac{r_{s}}{R}\right)
^{m}\right] \tag{A.5}\label{eqnA.5}\\
&  +\frac{3m^{2}(m-1)}{R^{3}}\left(  \frac{r_{s}}{R}\right)  ^{2m-3}%
\delta\left[  1-\left(  \frac{r_{s}}{R}\right)  ^{m}\right]  +\frac{m^{3}%
}{R^{3}}\left(  \frac{r_{s}}{R}\right)  ^{3m-3}\frac{\delta\left[  1-\left(
\frac{r_{s}}{R}\right)  ^{m}\right]  }{\left[  1-\left(  \frac{r_{s}}%
{R}\right)  ^{m}\right]  }\nonumber
\end{align}

\begin{align}
f^{(4)}\left[  r_{s}\right]   &  =-\frac{m(m-1)(m-2)(m-3)}{R^{4}}\left(
\frac{r_{s}}{R}\right)  ^{m-4}H\left[  1-\left(  \frac{r_{s}}{R}\right)
^{m}\right] \tag{A.6}\label{eqnA.6}\\
&  +\frac{m^{2}(m-1)(7m-11)}{R^{4}}\left(  \frac{r_{s}}{R}\right)
^{2m-4}\delta\left[  1-\left(  \frac{r_{s}}{R}\right)  ^{m}\right] \nonumber\\
&  +\frac{6m^{3}(m-1)}{R^{4}}\left(  \frac{r_{s}}{R}\right)  ^{3m-4}%
\frac{\delta\left[  1-\left(  \frac{r_{s}}{R}\right)  ^{m}\right]  }{\left[
1-\left(  \frac{r_{s}}{R}\right)  ^{m}\right]  }+\frac{2m^{4}}{R^{4}}\left(
\frac{r_{s}}{R}\right)  ^{4m-4}\frac{\delta\left[  1-\left(  \frac{r_{s}}%
{R}\right)  ^{m}\right]  }{\left[  1-\left(  \frac{r_{s}}{R}\right)
^{m}\right]  ^{2}}\nonumber
\end{align}

By inserting these derivatives into Eq. (\ref{eqnA.1}), we obtained the
expression used to compute $T^{00}$\ in Fig. 1 (bottom right panel), Fig. 2,
Fig. 4, and Fig. 6.

\begin{acknowledgments}
This work was supported by a grant from the Frank R. Seaver College of Science
and Engineering, Loyola Marymount University. The authors would like to
acknowledge suggestions and clarifications by Dr. P. Mannheim.
\end{acknowledgments}

\newpage

\bibliographystyle{apsrev}
\bibliography{CONFORMAL,COSMOBOOKS,MANNHEIM_RECENT,VARIESCHI,WARPDRIVE}
\newpage

\begin{center}
FIGURE CAPTIONS
\end{center}

Figure 1: Results for the two different shaping functions (left column ASF,
right column HSF), computed with parameters $v_{s}=1$, $R=1$, $\sigma=8$,
$m=4$, $\alpha_{g}=1$, $t=0$. Top row: Alcubierre and Hartle shaping
functions. Second row: expansion of the volume elements $\theta$. Third row:
energy density $T^{00}$ computed with General Relativity. Bottom row: energy
density $T^{00}$ computed with Conformal Gravity. The WEC is verified in the
case shown in the bottom right panel.

Figure 2: Energy density $T^{00}$ computed with Conformal Gravity and the
Hartle shaping function. Parameters used: $R=1$, $m=4$, $\alpha_{g}=1$, $t=0$,
and variable $v_{s}=0.25\ c-3.00\ c$. The energy density becomes in part
negative for $v_{s}\gtrsim2.50\ c$, so the WEC is verified for speeds up to
about $v_{s}\simeq2.50\ c$.

Figure 3: Stress-energy tensor components $T_{\mu\nu}$, in cylindrical
coordinates $(t,x,\rho,\phi)\equiv(0,1,2,3)$,\ computed with Conformal Gravity
and the Hartle shaping function. Parameters used: $v_{s}=1$, $R=1$, $m=4$,
$\alpha_{g}=1$, $t=0$.

Figure 4: Energy density $T^{00}$ computed with Conformal Gravity and the
Hartle shaping function. Parameters used: $v_{s}=1$, $R=1$, $\alpha_{g}=1$,
$t=0$, and variable $m=2-10$. In all cases the WEC is verified.

Figure 5: Violation of the DEC in the case analyzed (CG with HSF and $m=4$,
$R=1$, $v_{s}=1.00\ c$, $\alpha_{g}=1$, $t=0$). The plotted function
$T^{0}\,_{\lambda}T^{0\lambda}$ is not negative everywhere, as required by the
DEC, but shows a violation for the central portion of the warp bubble.

Figure 6: Energy density $T^{00}$ computed with Conformal Gravity and the
Hartle shaping function, plotted for $\rho\geq0$. Parameters used: $v_{s}=1$,
$R=10000$, $m=4$, $\alpha_{g}=1$, $t=0$. Integrating this local energy density
over all space, we obtain an estimate for the total energy $E$ required to
establish the warp drive.

\begin{figure}[ptb]%
\centering
\ifcase\msipdfoutput
\includegraphics[
width=\textwidth
]
{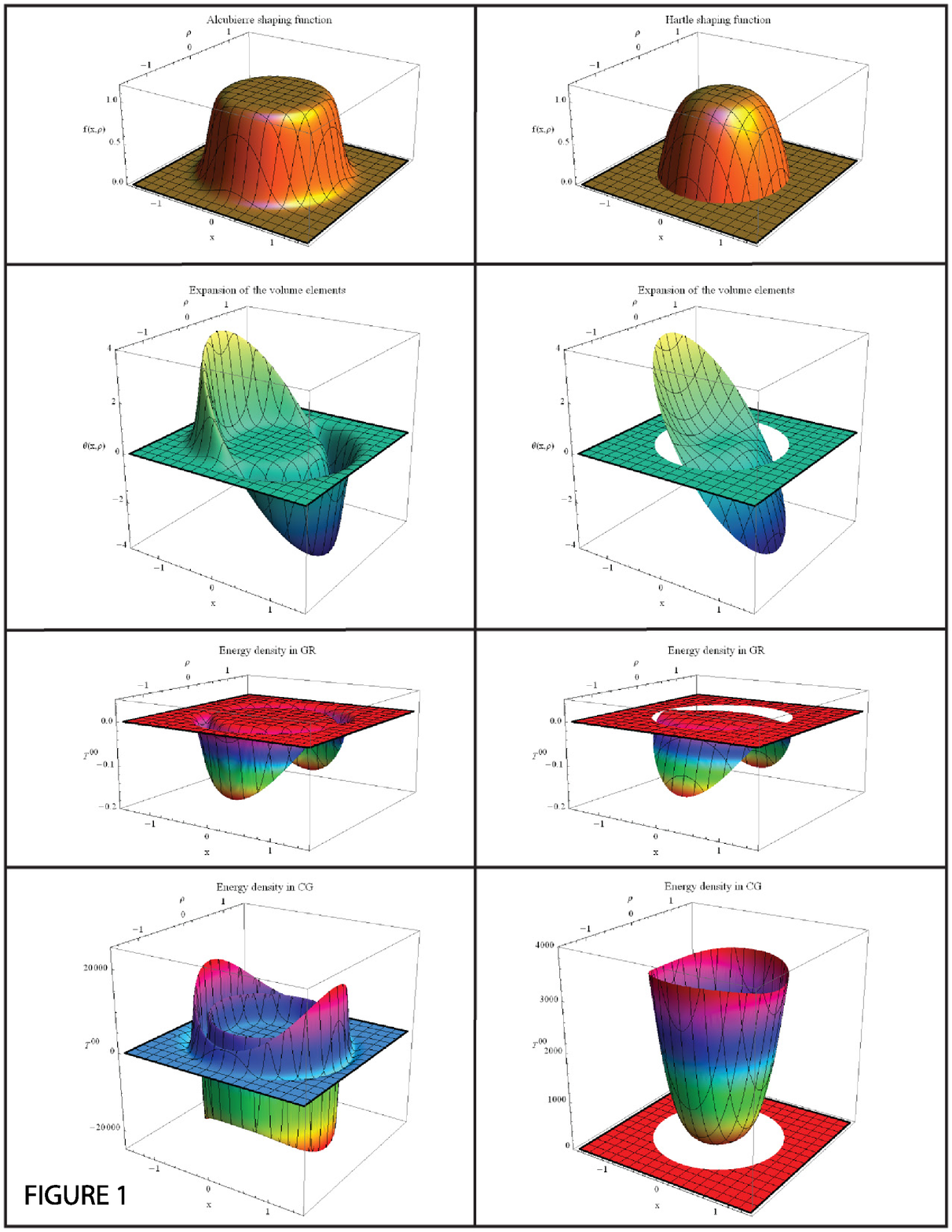}%
\else
\includegraphics[
width=\textwidth
]
{Varieschi_Burstein_fig1.eps}%
\fi
\label{fig1}%
\end{figure}

\begin{figure}[ptb]%
\centering
\ifcase\msipdfoutput
\includegraphics[
width=\textwidth
]
{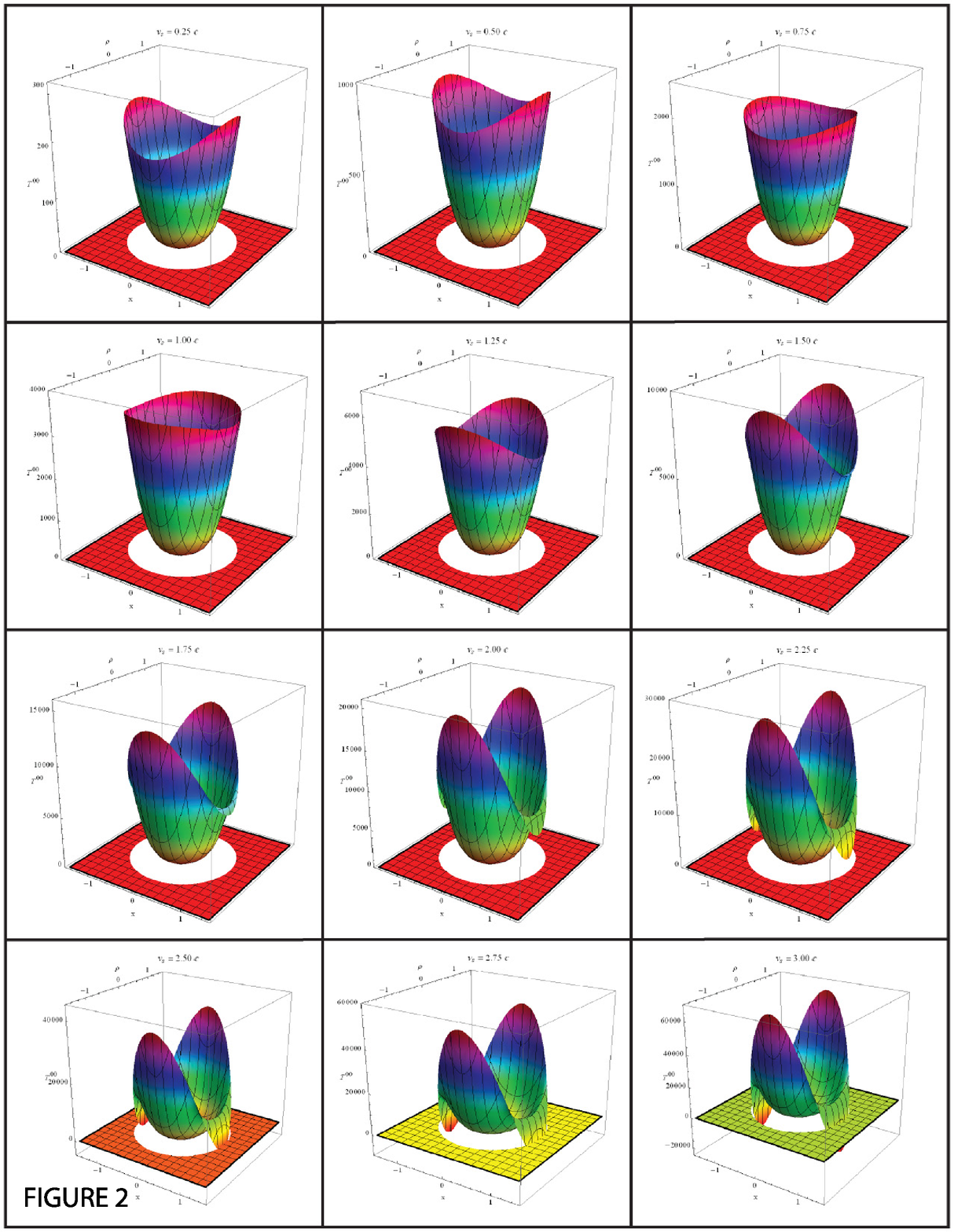}%
\else
\includegraphics[
width=\textwidth
]
{Varieschi_Burstein_fig2.eps}%
\fi
\label{fig2}%
\end{figure}

\begin{figure}[ptb]%
\centering
\ifcase\msipdfoutput
\includegraphics[
width=\textwidth
]
{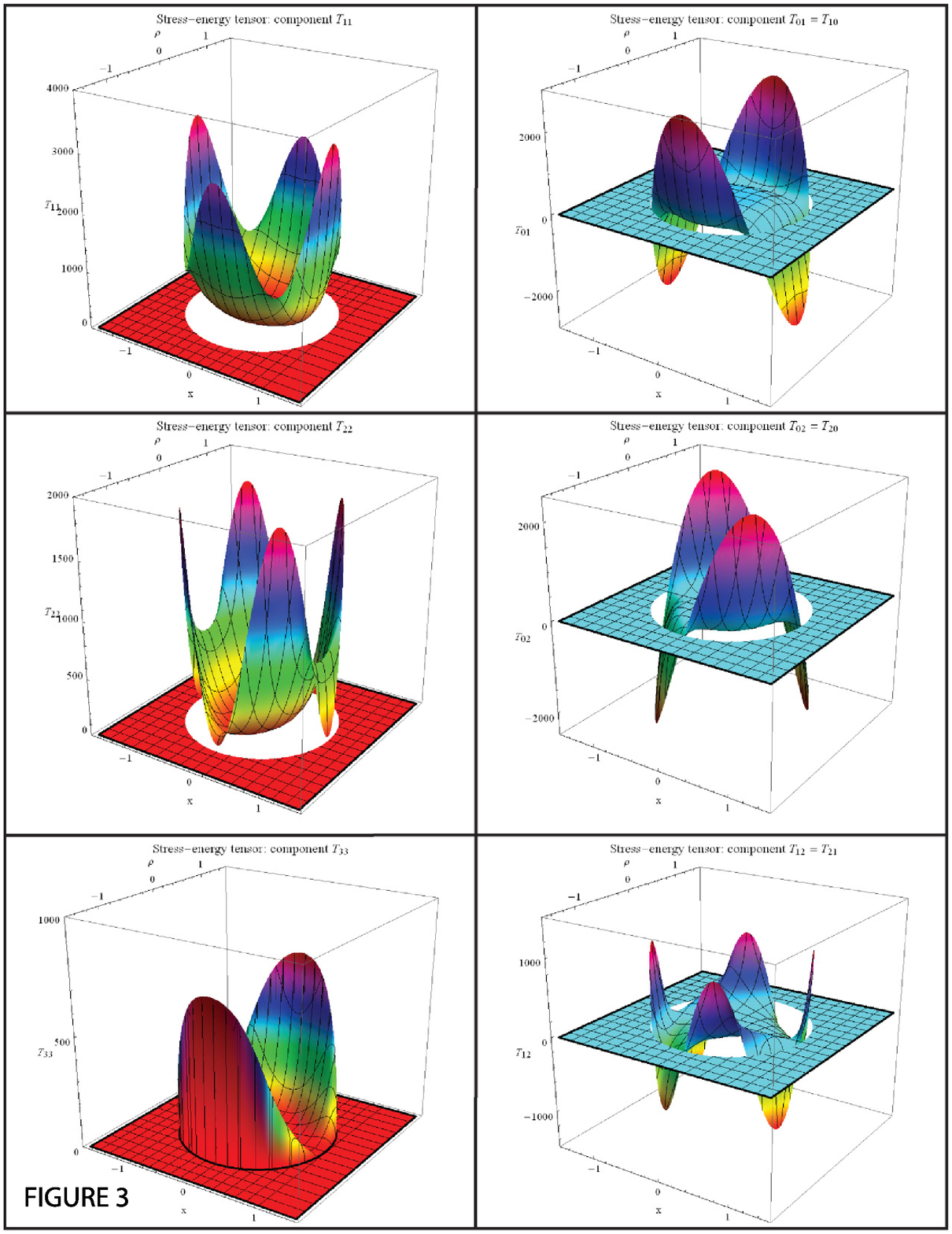}%
\else
\includegraphics[
width=\textwidth
]
{Varieschi_Burstein_fig3.eps}%
\fi
\label{fig3}%
\end{figure}

\begin{figure}[ptb]%
\centering
\ifcase\msipdfoutput
\includegraphics[
width=\textwidth
]
{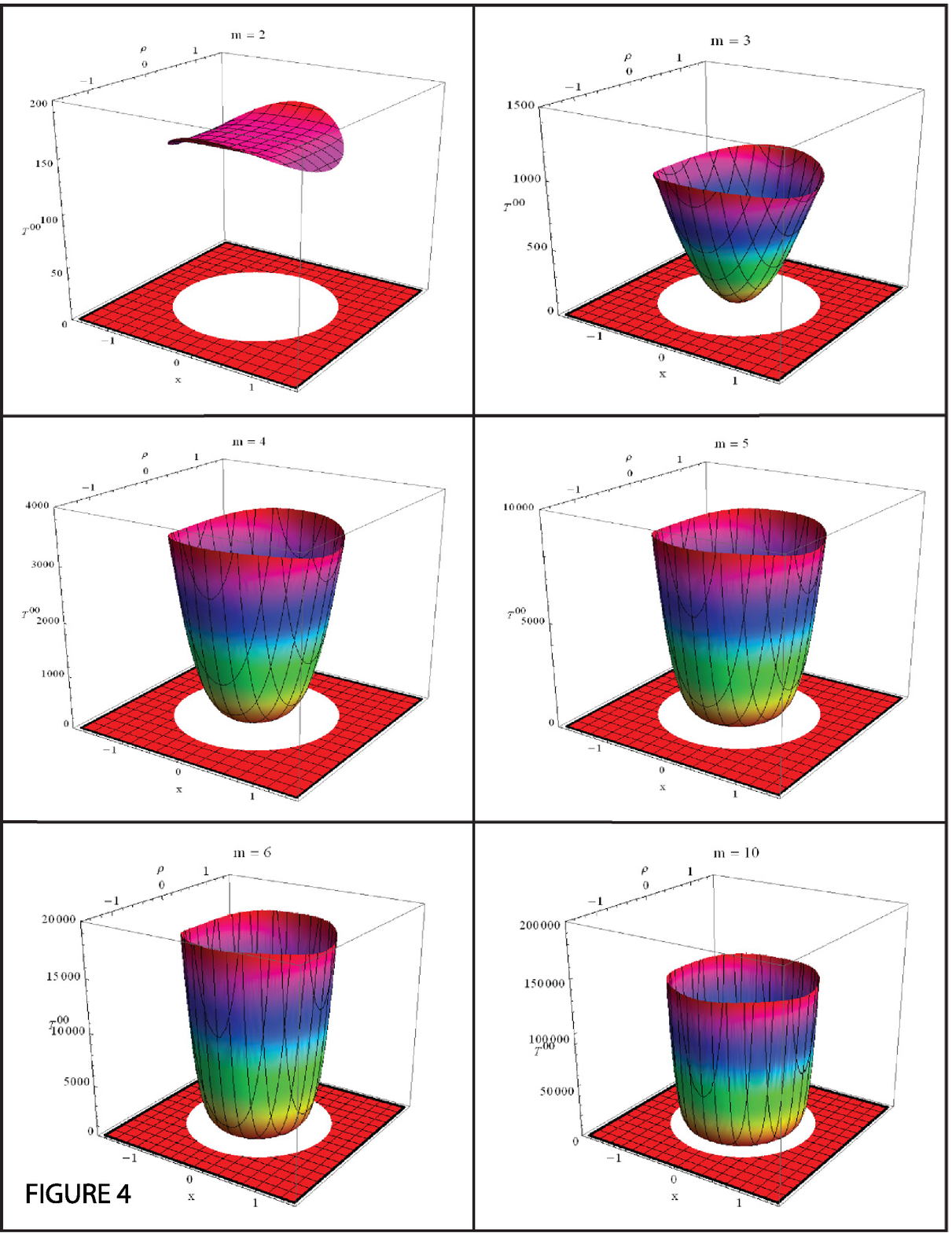}%
\else
\includegraphics[
width=\textwidth
]
{Varieschi_Burstein_fig4.eps}%
\fi
\label{fig4}%
\end{figure}

\begin{figure}[ptb]%
\centering
\ifcase\msipdfoutput
\includegraphics[
width=\textwidth
]
{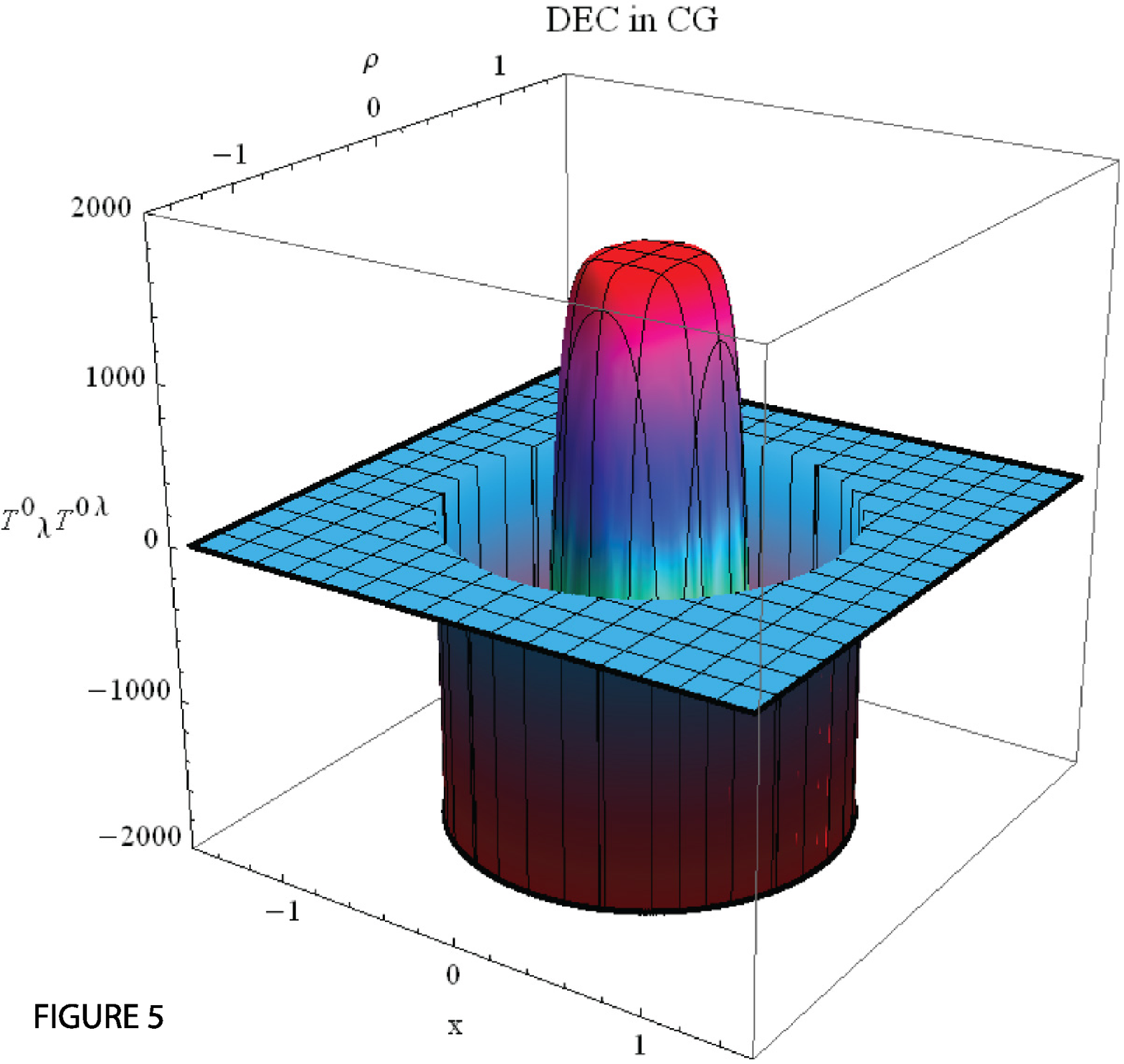}%
\else
\includegraphics[
width=\textwidth
]
{Varieschi_Burstein_fig5.eps}%
\fi
\label{fig5}%
\end{figure}

\begin{figure}[ptb]%
\centering
\ifcase\msipdfoutput
\includegraphics[
width=\textwidth
]
{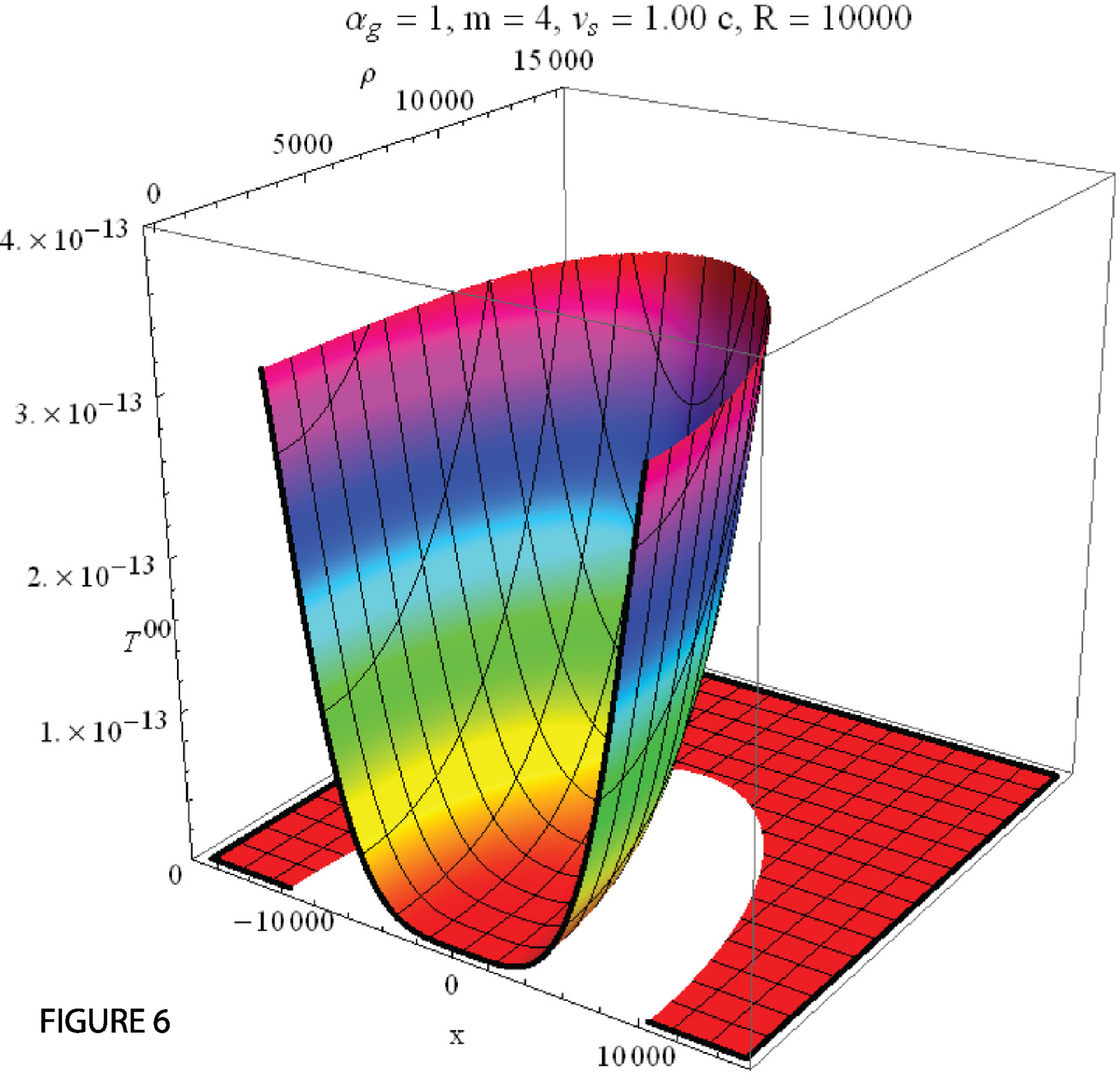}%
\else
\includegraphics[
width=\textwidth
]
{Varieschi_Burstein_fig6.eps}%
\fi
\label{fig6}%
\end{figure}

\end{document}